# Effect of α-particle irradiation on a NdFeAs(O,F) thin film

C Tarantini[1], K Iida[2,3], N Sumiya[2], M Chihara[2], T Hatano[2,3], H Ikuta[2,3], R K Singh[4], N Newman[4] and D C Larbalestier[1]

[1] National High Magnetic Field Laboratory, Florida State University, Tallahassee, Florida 32310, USA
[2] Department of Crystalline Materials Science, Nagoya University, Chikusa-ku, Nagoya 464-8603, Japan
[3] Department of Materials Physics, Nagoya University, Chikusa-ku, Nagoya 464-8603, Japan
[4] Materials Program, Arizona State University, Tempe, 85287 Arizona, USA

Email: tarantini@asc.magnet.fsu.edu

**Abstract**
The effect of α-particle irradiation on a NdFeAs(O,F) thin film has been investigated to determine how the introduction of defects affects basic superconducting properties, including the critical temperature $T_c$ and the upper critical field $H_{c2}$, and properties more of interest for applications, like the critical current density $J_c$ and the related pinning landscape. The irradiation-induced suppression of the film $T_c$ is significantly smaller than on a similarly damaged single crystal. Moreover $H_{c2}$ behaves differently, depending on the field orientation: for H//$c$ the $H_{c2}$ slope monotonically increases with increasing disorder, whereas for H//$ab$ it remains constant at low dose and it increases only when the sample is highly disordered. This suggests that a much higher damage level is necessary to drive the NdFeAs(O,F) thin film into the dirty limit. Despite the increase in the low temperature $H_{c2}$, the effects on the $J_c$(H//$c$) performances are moderate in the measured temperature and field ranges, with a shifting of the pinning force maximum from 4.5 T to 6 T after an irradiation of $2\times10^{15}$ cm$^{-2}$. On the contrary, $J_c$(H//$ab$) is always suppressed. The analysis demonstrates that irradiation does introduce point defects acting as pinning centres proportionally to the irradiation fluence but also suppresses the effectiveness of $c$-axis correlated pinning present in the pristine sample. We estimate that significant performance improvements may be possible at high field or at temperatures below 10 K. The suppression of the $J_c$(H//$ab$) performance is not related to a decrease of the $J_c$ anisotropy as found in other superconductors. Instead it is due to the presence of point defects that decrease the efficiency of the $ab$-plane intrinsic pinning typical of materials with a layered structure.

## 1. Introduction

Irradiation of superconductors has been performed for many years to improve our fundamental knowledge of the influence of impurity scattering on the superconducting properties and to explore whether the material performance can be improved for practical applications. For instance irradiation with neutrons, α-particles or electrons of A15 phases like V$_3$Si [1,2], Nb$_3$Al [3], V$_3$Ge [1], Nb$_3$Sn and Nb$_3$Ge [1,4,5] was employed to determine the effect on $T_c$ suppression, changes in the density of states (DOS) and in $H_{c2}$, variation in the gap amplitude and in the Debye temperature. More recently neutron and proton irradiation was found to favourably change the pinning landscape in order to increase both $H_{c2}$ and $J_c$ in commercial Nb$_3$Sn wires for accelerator magnets [6,7]. In the two-bands, two-gaps MgB$_2$ neutron irradiation was an effective tool to demonstrate the gap merging [8,9], which was theoretically predicted to occur with the $T_c$ suppression. Moreover neutron and α-particles irradiation also produces an improvement of $H_{c2}$ and $J_c$ in MgB$_2$ [10,11,12,13]. In the case of high-$T_c$ superconductors heavy ion, proton and electron irradiation was used to investigate many theoretical aspects like weak localization and Kondo-like behaviour [14] or to study the vortex phase diagram in YBa$_2$Cu$_3$O$_{7-\delta}$ (YBCO) [15,16]. The use of heavy ion and proton irradiations are also an efficient method to introduce strongly-correlated columnar or point defects which enhance the irreversibility field and $J_c$ at high field and high temperature in YBCO films [17,18] or coated conductors [19]. In Fe-based superconductors irradiation with neutrons, α-particles, proton and electrons was employed to investigate the unconventional pairing mechanisms and the gap symmetry in LaFeAs(O,F),

NdFeAs(O,F), Ba(Fe,Co)$_2$As$_2$ and Ba(Fe$_{1-x}$Ru$_x$)$_2$As$_2$ [20,21,22,23,24]. Different types of irradiation were performed on Fe-based superconductors with the intention of introducing pinning centres: columnar defects introduced by heavy ion irradiation were particularly effective in increasing $J_c$ in Ba(Fe,Co)$_2$As$_2$ [25] and (Ba,K)Fe$_2$As$_2$ [26]. Quite different is the effect of proton irradiation on Ba(Fe,Co)$_2$As$_2$ films with naturally grown correlated defects: $H_{c2}$ was found unchanged but a high-field low-temperature $J_c$ improvement was found in particular for H//$ab$ due to the competition between intrinsic columnar and irradiation-introduced point defects [27]. Neutron irradiation on NdFeAs(O,F) single crystal [28] and polycrystalline SmFeAs(O,F) [29] enhances $J_c$ in a wide temperature range, suppressing a fishtail effect [i.e. secondary peak in $J_c$(H)] observed before irradiation. Performance improvement was also introduced in Fe(Se,Te) films by neutron [30] and proton irradiation [31].

In this paper we will present the influence of α-particle irradiation on the superconducting properties of a high-quality NdFeAs(O,F) (Nd-1111) thin film. We will compare our results on the resistivity enhancement and $T_c$-suppression upon irradiation with the results of a similar experiment performed on a single crystal [21]. The unusual changes in $H_{c2}$ and the non-monotonic variation in the material intrinsic anisotropy are reported. The effect of irradiation in the modification of the pinning landscape is discussed in light of the introduction of new flux pinning centres, including their role in the unexpected suppression of the intrinsic pinning typical of moderate-to-highly-anisotropic superconducting materials.

## 2. Experimental details

The sample under investigation is a 90 nm-thick Nd-1111 thin film deposited by molecular beam epitaxy (MBE) on a MgO(100) substrate at Nagoya University [32,33]. The pristine sample critical temperature was estimated in $T_{c,90\%}$ ~ 49.0 K and the high crystalline quality of the sample were verified by X-ray diffraction using the methodology described in ref. 34. The sample was prepared for transport characterization with Pt contact pads and a 32 µm-wide/1 mm-long bridge fabricated by laser cutting. The resistivity and $I$-$V$ measurements were performed in a 16 T physical property measurement system (PPMS) over a wide temperature range down to 10 K and in variable field orientations, while maintaining a maximum Lorentz force configuration. $H_{c2}$ was evaluated with a 90% resistivity criterion, whereas $J_c$ was determined from the $I$-$V$ curves by a 1 µV/cm criterion. After the initial characterizations, the sample was irradiated in steps by α-particles at room temperature using a 2 MeV 4He$^{2+}$ ion beam from a Tandem accelerator at the Arizona State University with subsequent transport characterization measurements carried out at Florida State University. To avoid damage to the contacts, the sample was never dismounted from the transport-measurement sample holder. The sample was irradiated 3 times achieving the

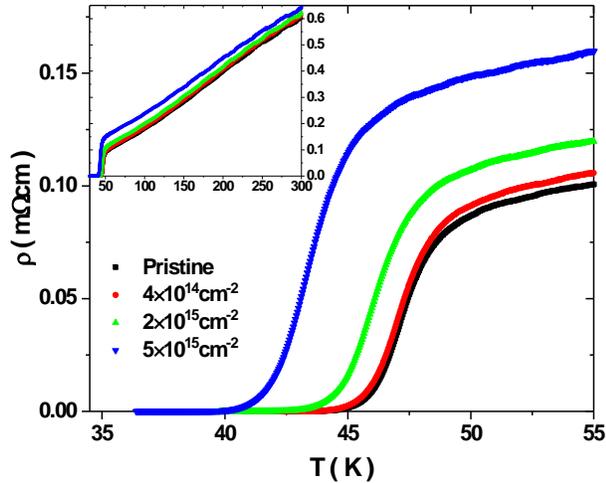

**Figure 1** Resistive transition of a NdFeAs(O,F) thin film before and after irradiation with α-particles. In the main panel, a magnification of the transitions is plotted whereas the inset shows the entire temperature range of measurement.

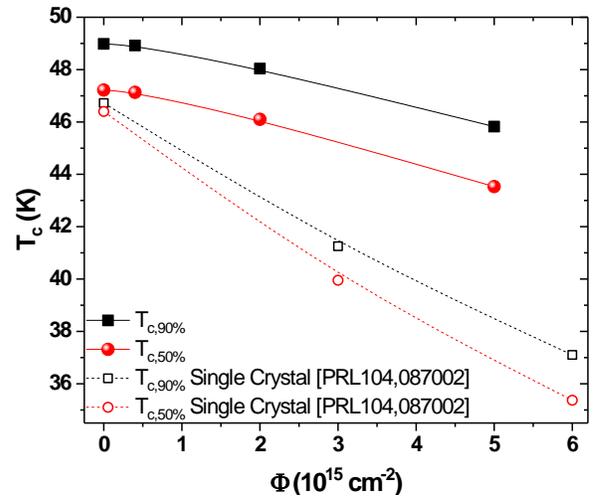

**Figure 2** $T_c$ versus fluence of α-particles for a NdFeAs(O,F) thin film. Reported trends were estimated at both 90% and 50%-criterion and compared with similar data on single crystal [21].

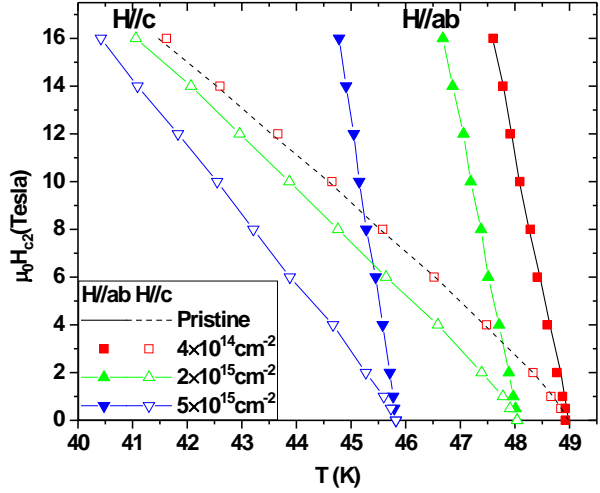

**Figure 3** Temperature dependence of $H_{c2}$ in the two main crystallographic configurations (H//*ab*, H//*c*) for a NdFeAs(O,F) thin film before and after irradiation with α-particles.

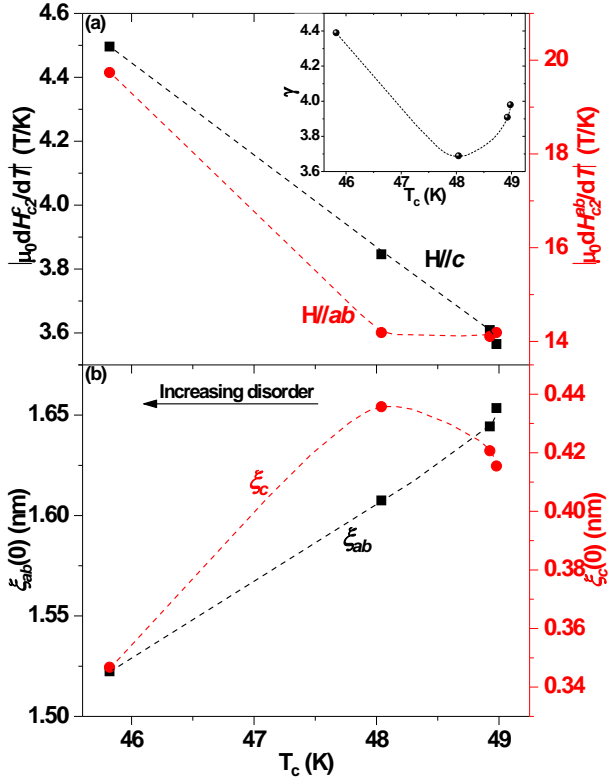

**Figure 4** (a) $H_{c2}$ slopes near $T_c$ calculated from the data in Figure 3 for the two main crystallographic orientations of a NdFeAs(O,F) thin film before and after irradiation with α-particles; in the inset the electronic mass anisotropy $\gamma$. (b) Coherence lengths at 0 K, $\xi_c(0)$ and $\xi_{ab}(0)$, estimated from the $H_{c2}$ slope near $T_c$.

following fluences, Φ: $4\times10^{14}$, $2\times10^{15}$ and $5\times10^{15}$ cm$^{-2}$. Since the estimated mean free path of α-particles in Nd-1111 is 4.2 µm [21], much larger than the film thickness, uniform irradiation damage is expected throughout the sample. Based on the changes found in the previous Nd-1111 single crystal irradiation experiment [21], the first irradiation dose was selected to achieve a $T_c$ suppression of less than 1 K.

## 3. Results

### 3.1. Resistive transition and $T_c$

The resistive transitions obtained before and after irradiation are reported in Figure 1. The pristine sample reveals a normal state resistivity, ρ(55K), of ~ 0.1 mΩcm, a sharp transition and resistivity ratio [RRR=ρ(300K)/ ρ(55K)] of ~ 6. Upon irradiation, ρ(55K) progressively increases and, after the last irradiation cycle, it is about 60% larger with a RRR reduced to 4. Irradiation produces only a modest $T_c$ suppression compared to a similarly irradiated single crystal (Figure 2) and mild transition broadening: $T_{c,90\%}$ changes by ~3.2 K at the maximum fluence, while $T_{c,50\%}$ changes by ~3.7 K. $\Delta T_c = T_{c,90\%} - T_{c,10\%}$ varies from ~ 3.1 K to ~3.9 K with irradiation.

### 3.2. $H_{c2}$ characterization and analysis

The upper critical field $H_{c2}$ was measured with the field aligned along the two main crystallographic directions before and after irradiation (Figure 3). After the first irradiation cycle, the $H_{c2}$ temperature dependence barely changed. At Φ = $2\times10^{15}$ cm$^{-2}$, $T_{c,90\%}$ was suppressed by about 1 K and $H_{c2}^{ab}$ appears simply shifted at lower temperatures, whereas $H_{c2}^c$ clearly shows an increase in its slope. At Φ = $5\times10^{15}$ cm$^{-2}$, $T_c$ is furtherly reduced and both $H_{c2}^{ab}$ and $H_{c2}^c$ reveal a slope increase. In order to quantify the intrinsic electronic mass anisotropy $\gamma$ and to estimate the coherence lengths, the $H_{c2}$ slopes were estimated near $T_c$ and plotted in Figure 4(a): with increasing disorder (decreasing $T_c$) a monotonic increase of $\left|\mu_0 \, dH_{c2}^c / dT\right|_{T_c}$ is observed while $\left|\mu_0 \, dH_{c2}^{ab} / dT\right|_{T_c}$ remains substantially unchanged up to the second irradiation step ($T_c$ ~ 48 K) and then it suddenly increases with further disorder. As a consequence, $\gamma$ has a non-monotonic trend with a minimum when $T_c$ is ~ 48 K. The coherence lengths at 0 K, $\xi_c(0)$ and $\xi_{ab}(0)$, can be estimated from the $H_{c2}$ slope at $T_c$ by combining the Ginzburg-Landau expressions with the Werthamer, Helfand and Hohenberg (WHH) formula [35, 36]

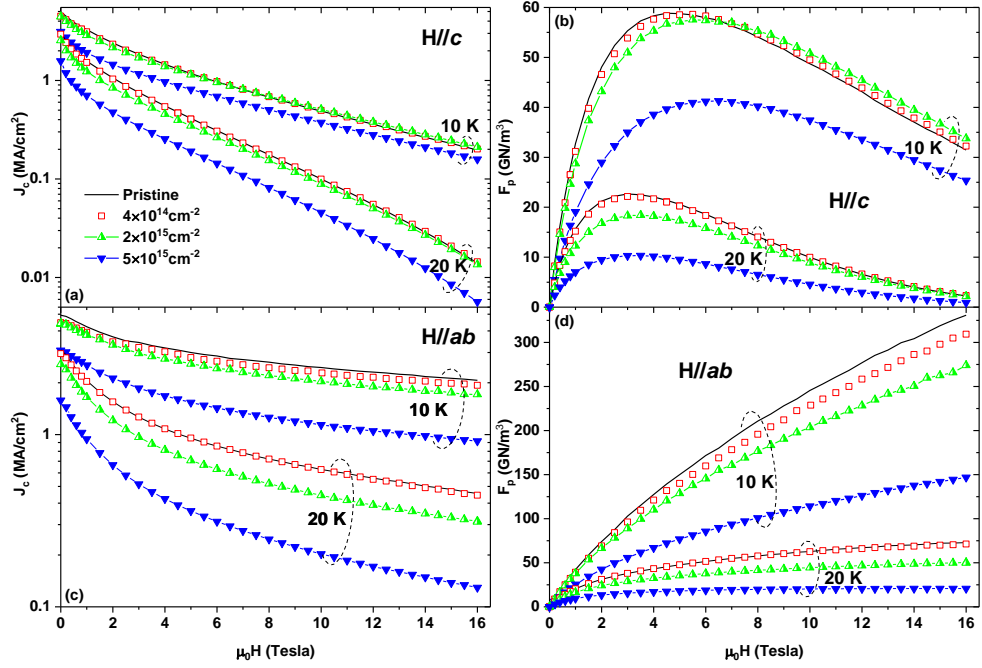

**Figure 5** Field dependences of $J_c$ and $F_p$ at 10 and 20 K for a NdFeAs(O,F) thin film before and after irradiation with α-particles for (a),(b) H//c and for (c),(d) H//ab.

$$\xi_{ab}(0) = [\phi_0 /(2\pi\mu_0 0.69 T_c |dH_{c2}^c/dT|_{T_c})]^{1/2}$$ and $\xi_c(0) = \xi_{ab}(0)/\gamma$. $\xi_{ab}(0)$ progressively decreases with increasing disorder, whereas $\xi_c(0)$ has a non-monotonic trend with a maximum at $T_c \sim 48$ K.

### 3.3. $J_c$ characterization and analysis

$J_c$ was measured from $T_c$ down to 10 K using 5 K increments at low temperatures and 2.5 K steps near $T_c$ for the as-grown sample and then after irradiation. Figure 5 shows the data at 10 and 20 K as an example. In the pristine sample the self-field $J_c$(10K) is ~ 5 MA/cm$^2$, whereas the $J_c$(16T,10K) values are just below 0.2 MA/cm$^2$ and more than 2 MA/cm$^2$ for H//c and H//ab, respectively. The pinning force density $F_p$(10K,H//c) reaches a maximum of 58.9 GN/m$^3$ at 4.5 T, whereas along the ab-planes the maximum exceeds 300 GN/m$^3$ which occurs above 16 T, outside the measurement range. After irradiation a small increase of $J_c$(H//c) at high field is observed for data below 25 K for $\Phi = 4\times10^{14}$ cm$^{-2}$ and a slightly larger one below 15 K for $\Phi = 2\times10^{15}$ cm$^{-2}$. More obvious is the change of $F_p$(H//c) shape with a shift of the maximum to 5.5-6 T for a fluence of $\Phi = 2\times10^{15}$ cm$^{-2}$. For H//ab $J_c$ and $F_p$ are progressively suppressed with increasing irradiation over substantially the entire temperature range.

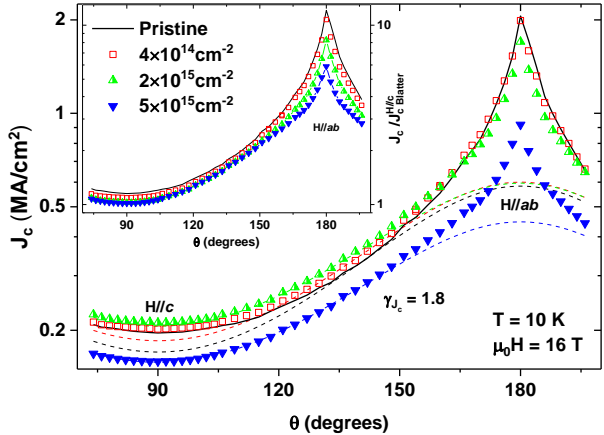

**Figure 6** Angular dependence of $J_c$ at 10 K and 16 T for a NdFeAs(O,F) thin film before and after irradiation with α-particles. The dashed lines represent the random pinning contributions as obtained from Blatter rescaling. In the inset, the same data normalized to $J_{c,\text{Blatter}}$(H//c).

Figure 6 reports a representative example of the angular dependence of $J_c$ at 10 K and 16 T: it shows that the largest increase in $J_c$ is reached after the second irradiation step in an angular range up to 165°. Note that as the field angle approaches that of the *ab*-planes, $J_c$ is suppressed by irradiation. After the third irradiation cycle $J_c$ is significantly suppressed over the entire angular range. The $J_c(\theta)$ curves, measured from 1 to 16 T, can be rescaled using the Blatter approach [37] with the same anisotropy parameter (~1.8) before and after irradiation [note: $J_c$ anisotropy differs from the intrinsic or $H_{c2}$ anisotropies in FBS as a consequence of the multiband-behaviour and the paramagnetic limit]. The dashed lines in Figure 6 represent the random pinning contribution, estimated using rescaling, whereas the inset presents the same data normalized to $J_{c,\text{Blatter}}$(H//$c$). These figures emphasize the effect of the correlated pinning contributions to $J_c$ as the field approaches the *c*-axis and the *ab*-planes.

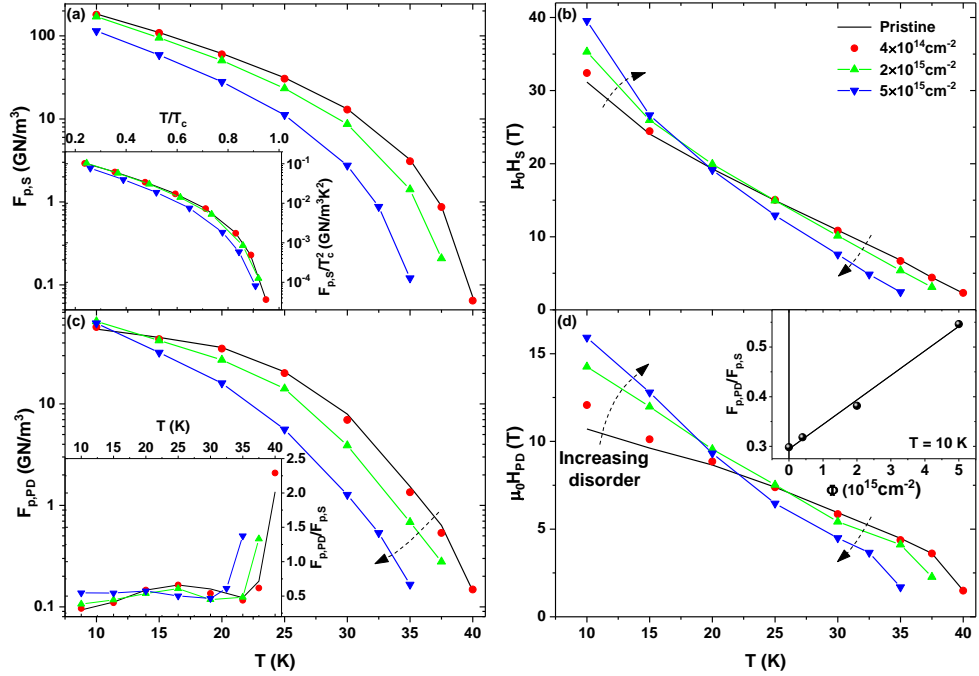

**Figure 7** Temperature dependence of the fitting parameters of the $F_p$(H//$c$) curves using $S$ [(a) and (b)] and $PD$ [(c) and (d)] contributions for a NdFeAs(O,F) thin film before and after irradiation with α-particles (see text for details). In the inset of panel (a) $F_{p,S}$ normalized to $T_c^2$ versus the reduced temperature T/$T_c$. In the inset of panel (c) the ratio between the $PD$ and the $S$ amplitude parameters, $F_{p,PD}/F_{p,S}$ versus T. In the inset of panel (d) $F_{p,PD}/F_{p,S}$ as a function of the fluence Φ at 10 K.

To better understand the pinning mechanisms contributing before and after irradiation, we analysed the $F_p$ curves for H//$c$ in a wide temperature range with the model introduced in ref. [34] taking into account two superimposed and independent contributions: a interface contribution (*S*) (previously called "surface" [38]) and a point defect (*PD*) one. The *S* contribution is determined by planar defects parallel to the *c*-axis, such as domains, antiphase/twin boundaries and dislocation arrays, whereas *PD* pinning is likely induced by atomic defects, such as intrinsic point defects present in the as-grown sample and the vacancies and interstitial defects artificially introduced by irradiation. This model has four free parameters: two of them define the *S* and *PD* contribution amplitudes ($F_{p,S}$ and $F_{p,PD}$) and the other two define their maximum fields of effectiveness ($H_S$ and $H_{PD}$, the highest of the two corresponding to the irreversibility field). The analysis reveals that $H_S$ and $H_{PD}$ have a cross-over with irradiation [Figure 7(b)-(d)]: at high temperature they decrease with increasing disorder, whereas at low temperature they both show a clear increase. This is particularly obvious for $H_{PD}$, which increases by ~49% at 10 K (~27% for $H_S$). However the amplitudes of $F_{p,S}$ and $F_{p,PD}$ have a different trend [Figure 7(a)-(c)]. $F_{p,S}$ is systematically suppressed with irradiation in the investigated temperature range, whereas $F_{p,PD}$ is clearly suppressed at high temperature but at 10 K it increases up to 19% at $\Phi= 2 \times 10^{15}$ cm$^{-2}$. This, combined with the increase in $H_S$ and $H_{PD}$, can lead to a significantly improved performance at larger fields, with $J_c$ extrapolated from the fits improving by 27% at 20 T and almost doubling at 25 T ($\Phi= 2 \times 10^{15}$ cm$^{-2}$). Moreover, extrapolating to lower temperatures, we estimate that a noticeable improvement in the pinning force can be achieved also at low field: in fact at 2 K a maximum $F_p$ of 126 GN/m$^3$ at about 7 T for the pristine sample and 141 GN/m$^3$ at 8.3 T for a fluence of $\Phi = 2\times10^{15}$ cm$^{-2}$ are expected. The meaning of the insets of Figure 7 will be explained in the discussion section.

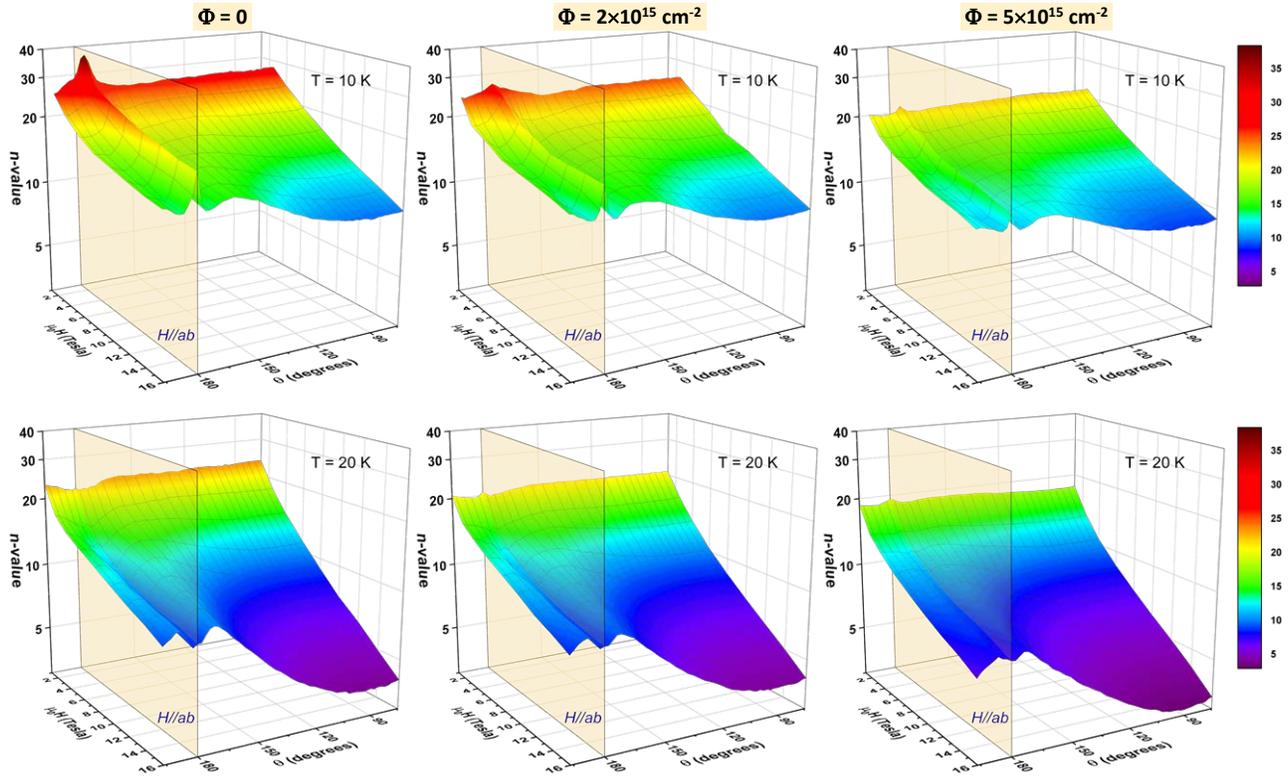

**Figure 8** Angular and field dependence of the *n*-values at 10 (1$^{st}$ row) and 20 K (2$^{nd}$ row) for a NdFeAs(O,F) thin film before irradiation (left) and after irradiation at $\Phi = 2\times10^{15}$ cm$^{-2}$ (middle) and $\Phi = 5\times10^{15}$ cm$^{-2}$ (right).

Since the suppression of $J_c$(H//$ab$) with irradiation cannot be ascribed to a change in the $J_c$ anisotropy (Blatter rescaling is performed with the same value), we analyse the *n*-values to better understand the role of irradiation on the film properties when the field approaches the $ab$-planes. The *n*-values are estimated from the *I-V* characteristics that are well described by the $V \propto I^n$ power-law. Figure 8 shows exemplary $n(\theta,H)$ curves at 10 and 20 K for the pristine sample and after irradiations at $2 \times 10^{15}$ and $5\times10^{15}$ cm$^{-2}$. At high temperature (not shown), the $n(\theta)$ curves have a minimum at H//$c$ and a maximum at H//$ab$, similar to the angular dependence of $J_c$ and the data follow the expected $n \sim J_c^\alpha$ relation [39,34]. At 30-25 K, the sharpness of the $ab$-peak is suppressed and, at 25-20 K, a dip at H//$ab$ appears due to vortex trapping by the Nd-1111 layers and the consequent formation of a vortex-staircase structure [34,35,40,41]. The data at 20 K in Figure 8 show this dip (both before and after irradiation) with a flat valley bottom at high field and a small peak emerging in the middle at low field. At 10 K the dip widens and a peak more clearly emerges due to the vortices being locked-in between the layers along the entire sample width

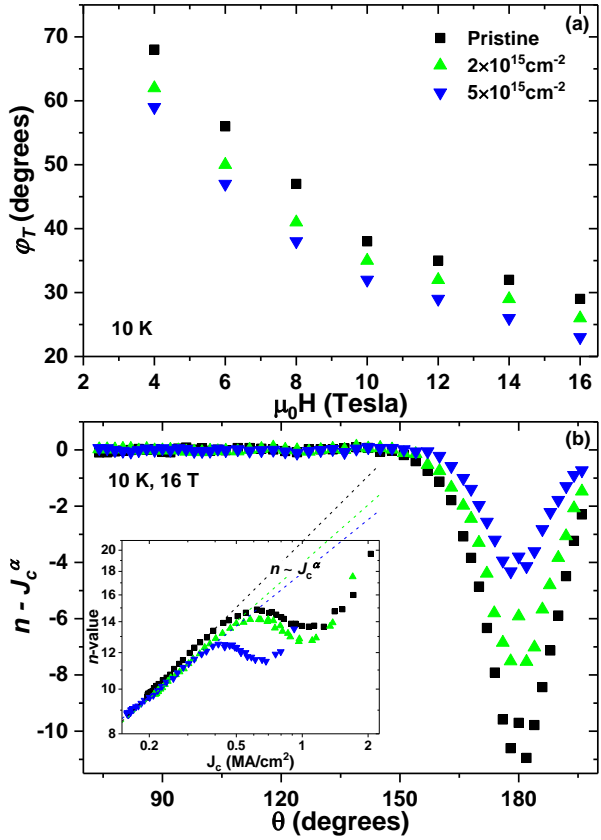

**Figure 9** (a) Trapping angles as a function of applied field at 10 K for a NdFeAs(O,F) thin film before and after irradiation with α-particles. (b) Angular dependence of $n$-$J_c^\alpha$ obtained by subtracting the power-law behaviour found in the $n(J_c)$ plot (see inset) from the $n$-value (at 10 K and 16 T).

[34,35,40,41]. After irradiation, there is a global lowering of the $n$-value and the dip depth and its angular range appears different. To quantify these differences, we estimated the trapping angle $\varphi_T(10\,K, H)$ as the angle from the $ab$-planes at which the data deviate from $n \sim J_c^\alpha$ [34]. Figure 9 (a) shows that the $\varphi_T$ curves roughly follow the same field dependence, but $\varphi_T$ progressively decreases with disorder. Subtracting the $n \sim J_c^\alpha$ trend from the $n(\theta)$ curve emphasizes the variation in the dip depth: the data in Figure 9 (b) and its inset report an example at 10 K and 16 T and it reveals that the dip become more shallow with increasing disorder.

## 4. Discussion

The normal state behaviour and the superconducting transition observed in the Nd-1111 thin film are very different with respect to what we previously observed on single crystal Nd-1111 [21]. In fact, the latter showed a progressive increase in the normal state resistivity with a clear and significant upturn at low temperatures that was already present after the first irradiation step ($\Phi = 3 \times 10^{15}$ cm$^{-2}$). In contrast, our film shows only a small change in the shape of ρ(T) after a similar irradiation. Also the suppression of $T_c$ by the irradiation is significantly smaller in the film than the single crystal by a factor of ~2.5 (Figure 2). The reason for these unexpected differences is uncertain. Since the atom displacements mainly occur on the Nd, Fe and As sites [21] (with the first two elements being magnetic), the upturn observed in the single crystal was ascribed to Kondo-like magnetic scattering induced by irradiation defects [21]. A possible explanation for the unexpected differences of the post-irradiation properties could be a result of their different doping levels. Since the underdoped single crystal is closer to the antiferromagnetic phase characteristic of Fe-based superconductors [42,43], it may be more susceptible to magnetic disorder induced by irradiation defects and thus it could induce local off-stoichiometric regions, allowing antiferromagnetism to re-emerge at a much lower level of disorder. Another possible explanation is that in the film the substrate presence and the induced strain prevent the structure from fully relaxing after an irradiation displacement occurs; this could facilitate an increased rate of interstitial-vacancy recombination, partially recovering the disorder.

In regard to $H_{c2}$ and the coherence lengths, we observe different behaviours depending on the field orientations. For H//$c$ the $H_{c2}$ slope monotonically increases (and $\xi_{ab}(0)$ decreases) with irradiation, implying that the dirty limit regime is immediately established (the dirty limit is reached when the mean free path is comparable with the coherence length [44]). For H//$ab$, the slope is unchanged after the first two irradiation cycles (for $T_c \geq 48$ K) and then it suddenly increases after the third cycle. This means that a larger level of disorder is necessary to establish the dirty limit in this orientation: this is not surprising considering that the $c$-axis coherence length is about 4 times smaller than along *ab*. These different coherent length behaviours generate an unusual trend of γ (figure 4 inset) that in other materials typically decrease with increasing disorder [45,46].

The $J_c$ and $F_p$ performances of the pristine sample are significantly better, by a factor 2-3, than the previously reported results [34]. What is surprising for the present thin film is the ineffectiveness of the new pinning centres in increasing $J_c$ after irradiation. Even more, comparing $J_c(\theta)$ before and after irradiation with their respective random contributions (Figure 6) reveals that the *c*-axis correlated pinning present in the pristine sample almost disappears at $\Phi = 2 \times 10^{15}$ cm$^{-2}$. A possible reason for the reduction of the *c*-axis correlated pinning is that segregation of point defects (referred as Cottrell clouds) near planar defects is often found after irradiation as a result of the presence of both compressive and tensile strain in the region and the presence of some dangling bonds.

The two-contribution pinning model allows us to better understand the $J_c$ and $F_p$ behaviours. The decrease of the $H_S$ and $H_{PD}$ parameters at high temperature with increasing disorder (Figure 7) is clearly related to the $T_c$ suppression, whereas their low-temperature increase is due to the enhancement of the low-temperature $H_{c2}$ extrapolation, as inferred from Figure 3, and to the increase of point defect density in the case of $H_{PD}$. The amplitudes $F_{p,S}$ and $F_{p,PD}$ in general can be expressed it terms of fundamental and microstructure-related parameters as follows [38,47]:

$$F_{p,S} = \mu_0 S_v H_c^2/2 = S_v E_c \qquad (1)$$
$$F_{p,PD} = \mu_0 V_t H_c^2/2.32a = V_t E_c/1.16a \qquad (2)$$

where $\mu_0$ is the vacuum magnetic permittivity, $H_c$ is the thermodynamic critical field, $S_v$ is the pinning surface area per unit volume projected in the direction of the Lorentz force, $a$ is the average effective diameter of point defects, $V_t$ is the fraction of flux-lines length inside the pinning centre and $E_c = \mu_0 H_c^2/2$ is the condensation energy that is proportional to $N^* T_c^2$ (with $N^*$ being the renormalized electronic DOS) [47]. Since the microstructure-related parameter $S_v$ for planar defects does not change with irradiation, $F_{p,S}$ in eq. (1) should rescale with $T_c^2$ if $N^*$ remains constant. The inset of Figure 7(a) reports the $F_{p,S}$ data normalized to $T_c^2$ as a function of the reduced temperature $T/T_c$; it reveals that the curves obtained at $4\times10^{14}$ and $2\times10^{15}$ cm$^{-2}$ fluences follow the same trend as the pristine sample, whereas the $5\times10^{15}$ cm$^{-2}$ curve clearly falls below the others. This means that up to $\Phi = 2 \times 10^{15}$ cm$^{-2}$, $F_{p,S}$ only changes because of the $T_c$ suppression but after the last irradiation cycle also the DOS is reduced, provoking a significant drop in $J_c$. A similar rescaling does not work for the *PD* contribution (not shown): although, taking into account eq. (2), this is expected because the increasing point defect density with irradiation changes $V_t$, we also observe a change in the curve temperature dependences. The inset of Figure 7(c) helps to better understand the $F_{p,PD}$ change of trend: it shows the temperature dependence of the amplitude ratio, $F_{p,PD}/F_{p,S}$, that is independent from the condensation energy and equal to $V_t/1.16aS_v$. All curves have a maximum approaching $T_c$, a sort of plateau at intermediate temperature, but they separate at the lowest temperatures. In fact $F_{p,PD}/F_{p,S}(10K)$ almost doubles after the final cycle with respect to the pristine sample. If $L$ is the average distance between pinning centres, $V_t$ can vary from $(a/L)^3$ in the case of rigid vortices to $(a/L)$ in the case of perfectly flexible vortices or when $L \sim d$, with $d$ being the vortex spacing (matching effect)[38]; $F_{p,PD}/F_{p,S}$ assumes values between:

$$F_{p,PD}/F_{p,S} = a^2/1.16 S_v L^3 = 1/1.16 S_v L \cdot (a/L)^2 \qquad \text{for rigid vortices} \qquad (3)$$
and $$F_{p,PD}/F_{p,S} = 1/1.16 S_v L \qquad \text{perfectly flexible vortices (or } L \sim d\text{)} \qquad (4)$$

To verify these relation in the inset of Figure 7(d), we plot the $F_{p,PD}/F_{p,S}$ as a function of the fluence $\Phi$, which is proportional to $(1/L)^3$, at 10 K: at low temperature all point defects (either pre-existing or induced by irradiation) should be effective pinning centres. $F_{p,PD}/F_{p,S}$ versus $\Phi$ shows a roughly linear trend indicating that at this temperature $F_{p,PD}/F_{p,S}$ follows relation (3) and the vortices are rigid [$V_t \sim (a/L)^3$], as predicted by Dew-Hughes [38]. We can also note that, at $\Phi = 0$, $F_{p,PD}/F_{p,S}$ does not go to 0 as a result of the presence of pre-existing defects. When approaching $T_c$ [Figure 7(c) inset], $F_{p,PD}/F_{p,S}$ increases toward the limiting value in eq.(4) as a result of the increased flexibility of the vortex and the smaller applied fields, that are close to the matching field. At intermediate temperatures, the small variations in $F_{p,PD}/F_{p,S}$ are probably due to some point defects gradually losing effectiveness with increasing $\xi(T)$ and, as a consequence, changing the $a/L$ ratio in eq.(3).

The $n$-value carries information about the pinning performance properties through the relation $n \sim U_p/k_BT$ (with $U_p$ being the pinning potential and $k_B$ the Boltzmann constant) [35,48]. In the pristine sample at low temperature, when the vortex-staircase forms, the segments of vortices (kinks) connecting the trapped parts are not or are weakly pinned and they can easily move due to the Lorentz force being almost parallel to the $ab$-planes [40,34]: this leads to the $n$-value suppression and to the formation of the $n(\theta)$ dip when the field orientation approaches the $ab$-planes. When the field orientation is even closer to the $ab$-planes, the length of the trapped segments of the vortices becomes comparable with the sample width and the vortices become fully locked parallel to the $ab$-planes, generating the sudden increase in the $n$-value and the formation of the peak inside the dip. Clearly a similar effect occurs also after irradiation, but the progressive decrease of $\varphi_T$ and the dip shallowness with increasing irradiation are surprising. $\varphi_T$ is related to $\gamma$ (and so $\xi_c$) and the monotonic trend of $\varphi_T$ with disorder is inconsistent with the non-monotonic behaviour of $\gamma$ shown in Figure 4. The small increase of $\xi_c$ at $\Phi = 2\times10^{15}$ cm$^{-2}$ leads to a small decrease of $\varphi_T$ (as expected) but the following strong $\xi_c$ suppression at $\Phi = 5\times10^{15}$ cm$^{-2}$ should produce a clear increase of $\varphi_T$ due to the vortices being trapped over a wider angle around the $ab$-planes. The reason this increase of $\varphi_T$ does not occur and the dip becomes less pronounced is likely related to the increased density of point defects. In fact, the point defects introduced by irradiation are likely able to partially prevent the movement of the vortex kinks: as a consequence, although the staircase-structure forms, the vortex movement is inhibited until the field is aligned closer to the $ab$-planes (smaller $\varphi_T$ with increasing disorder) and the $n$-value dip is suppressed due to the additional pinning centres (smaller deviation from the $n \sim J_c^\alpha$ trend with increasing disorder). Moreover it is possible that pinning by point defects competes with the intrinsic pinning produced by the Nd-1111 layered structure, preventing the development of long vortex lengths lying parallel to the $ab$-planes and reducing the effect of intrinsic pinning in the formation of a sharp $J_c$ $ab$-peak.

## 5. Conclusions

Here we show that, despite the rather limited $T_c$ suppression upon α-particle irradiation of Nd-1111, $H_{c2}$ and $H_{irr}$ can be significantly enhanced by radiation-induced disorder. In contrast to the typically found monotonic irradiation-induced decrease in the electronic anisotropy found in most superconductors, we found an initial decrease of γ with increasing disorder, then followed by a sharp increase. The effects of irradiation disorder on $J_c$ depend on the field orientation. Point defect flux-pinning centres are indeed introduced in the system proportionally to α-particle fluence. However irradiation actually suppress the $c$-axis correlated pinning contribution by suppressing $T_c$ and by reducing the density of states. When the applied magnetic field approaches the $ab$-planes, the $J_c$ is reduced over the whole temperature range as a result of the competition between the point defect and intrinsic pinning mechanisms. Our results indicate that irradiation can improve the performance of Nd-1111 at high field and lower temperature and that it also can reduce the effective $J_c$ anisotropy.

**Acknowledgments**
A portion of this work was performed at the National High Magnetic Field Laboratory, which is supported by National Science Foundation Cooperative Agreement No. DMR-1157490 and the State of Florida. This research has been also supported by Strategic International Collaborative Research Program (SICORP), Japan Science and Technology Agency. K.I. and H.I. acknowledge support by the Japan Society for the Promotion of Science (JSPS) Grant-in-Aid for Scientific Research (B) Grant Number 16H04646.